\documentclass[aps,showpacs,twocolumn,nofootinbib]{revtex4}
\usepackage{graphics} 
\usepackage{graphicx} 
\usepackage{epstopdf}
\usepackage{color}
\usepackage{scrtime}
\usepackage{amssymb}
\usepackage{amsmath}
\usepackage{amsfonts}
\usepackage{hyperref}
\newcommand{\bra}[1]{\langle #1|}
\newcommand{\ket}[1]{|#1\rangle}

\begin{document}
\title{The Giant Pairing Vibration in Heavy Nuclei: Present Status and Future Studies }
\author{ M. Assi\'e}
\affiliation{ Institut de Physique Nucl\'eaire, CNRS/IN2P3, Universit\'e de Paris Sud, Universit\'e de Paris-Saclay, 91406 Orsay, France}
\author {C. H. Dasso}
\affiliation{ Departamento de Fisica Atomica, Molecular y Nuclear, Facultad de Fisica, Apartado 1065,
E-41080 Sevilla, Spain  }
 \author{R. J. Liotta}
\affiliation{KTH (Royal Institute of Technology), Alba Nova University Center, S-10691 Stockholm, Sweden}
 \author{  A.O. Macchiavelli}
 \affiliation{ Nuclear Science Division, Lawrence Berkeley National Laboratory, Berkeley CA 94720, USA}
 \author{A. Vitturi }
\affiliation{ Universita degli Studi di Padova and INFN, I-35131 Padova, Italy  }

%, and A. Vitturi \inst{5}

%}                     % Do not remove


\date{\today}

\begin{abstract}
The Giant Pairing Vibration, a two-nucleon collective mode originating from the second shell above the Fermi surface, has long been predicted and expected to be strongly populated in two-nucleon transfer reactions with cross sections similar to those of the normal Pairing Vibration.  Recent experiments have provided evidence for this mode in $^{14,15}$C but, despite sensitive studies, it has not been definitively identified either in Sn or Pb nuclei where pairing correlations are known to play a crucial role near their ground states.  In this paper we review the basic theoretical concepts of this "elusive" state and the status of experimental searches in heavy nuclei. We   discuss the hindrance effects due to Q-value mismatch and the use of weakly-bound projectiles as a way to overcome the limitations of the (p,t) and (t,p) reactions. 
We also discuss the role of the continuum and conclude with some possible future developments.
\end{abstract}

 %     {PACS-key}{discribing text of that key}   \and
  %    {PACS-key}{discribing text of that key}
 %    } % end of PACS codes
 %end of abstract
%

% \authorrunning{Assi\'e, Dasso, Liotta, Macchiavelli, Vitturi } 
 %\titlerunning {The GPV in Heavy Nuclei}
 
\maketitle

Pier Francesco Bortignon was a Giant in the studies of Giant Collective Modes in Atomic Nuclei.  His seminal works in this
and many other aspects of nuclear structure leave a strong legacy in the field,  and provide a guiding light for future developments. 
We are honored to  contribute to this special issue of EPJA, in memory of Pier Francesco, on a subject of much interest to him.
In fact, the last paper he wrote in collaboration with Ricardo Broglia  was entitled: {\sl Elastic response of the atomic nucleus in gauge space: Giant Pairing Vibrations}~\cite{PFB} and focused on the implications of the (rather unexpected) discovery of the Giant Pairing Vibration in light nuclei 
to further our knowledge of the basic mechanisms - Landau, doorway, and compound
damping- through which giant resonances acquire a finite lifetime.

This paper is dedicated to the memory of Pier Francesco and Hugo Sofia, another friend and
colleague who left us too early.  They are sorely missed.

\section{Introduction }
\label{intro}
Pairing correlations provided a key to understanding the excitation spectra of even-A nuclei,
odd-even mass differences, rotational moments of inertia, and a variety of other
phenomena~\cite{BMP,Bel,BB}. An early approach to describing pair correlations in nuclei was the
derivation of a collective Hamiltonian by B\`es and co-workers in formal analogy to the
Bohr collective Hamiltonian which describes the quadrupole degree of freedom for the
nuclear shape~\cite{Bes1}. The analogy between particle-hole (shape) and particle-particle
(pairing) excitations became well established and thoroughly explored by Broglia
and co-workers~\cite{BHR}, and more recently simple analytical approximations to the pairing collective
hamiltonian were used to describe the transition from normal to
superfluid behavior~\cite{CMFK}.

Nuclei with two identical particles added or removed from a closed-shell configuration
should be close to a  normal fluid limit, where there is no static deformation of the pair
field and fluctuations give rise to a pairing vibrational spectrum~\cite{BrogliaBes2}. Low-lying
pair-vibrational structures have been observed around $^{208}$Pb by using conventional
pair-transfer reactions such as  (p,t) and (t,p)~\cite{BHR,BM}. Nuclei with many particles outside of a
closed-shell configuration correspond to a superconducting limit, where there is a static
deformation of the pair field and rotational behavior results. An example would be the
pair-rotational sequence comprising the ground states of the even-even Sn isotopes
around $^{116}$Sn~\cite{BTG}.

It has long been predicted that there should be a concentration of strength, with L=0 character, in the high-energy region (10-15 MeV) of the 
pair-transfer spectrum.  This is called the Giant Pairing Vibration (GPV) and is understood 
microscopically  as the coherent superposition of 2p (addition mode) or 2h (removal mode)  states in the next major shell 2$\hbar\omega$
above (below) the Fermi surface ~\cite{BrogliaBes}.  Similar to the well-known pairing vibrational mode (PV)~\cite{BHR,BrogliaBes2,BM} which involves spin-zero-coupled pair excitations across a single major shell gap.

\begin{figure}

\resizebox{0.4\textwidth}{!}{\includegraphics{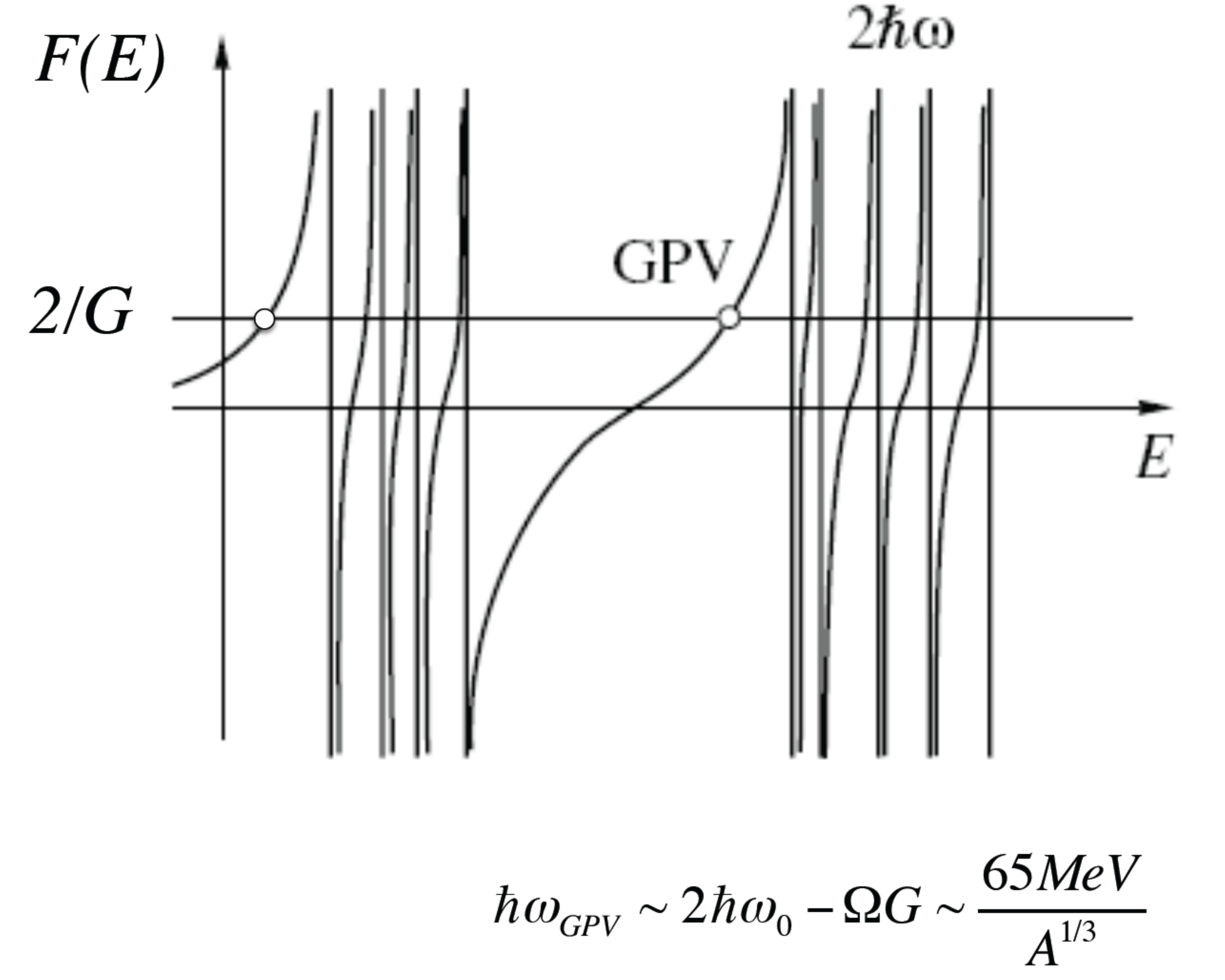}}
\caption{ Schematic of the dispersion relation, Eq. 2, showing the appearance of the collective GPV state and its estimated energy. The lowest solution corresponds to the PV.}
\label{Fig:fig1}
\end{figure}

Consider an schematic  Hamiltonian describing the motion of independent particles interacting by a (constant) pairing
force:

\begin{eqnarray}
H=  \sum_{j}  e_j ( a_{j}^\dagger a_{j}  + a_{\bar{j}}^\dagger a_{\bar{j} }  ) - G  \sum_{j,k}  a_{j}^\dagger a_{\bar{j}}^\dagger a_{k}a_{\bar{k}}
\end{eqnarray}
%\begin{align}
%P^\dagger = \sum_{j} a_{j}^\dagger a_{\bar{j}}^\dagger \nonumber
%end{align}

\noindent where the single-particle energies, measured from the Fermi surface, are denoted
by $e_{j}$, and the single-particle creation operators by $a_j^\dagger$.
The nature of the GPV is schematically illustrated in Fig. \ref{Fig:fig1},  that shows the solution of the dispersion relation obtained from a Random Phase Approximation (RPA) of the Hamiltonian in Eq. 1 \cite{BrogliaBes}
\begin{eqnarray}
 F(E)= \sum_{j}\frac{(2j+1)}{E-2e_j}=\frac{2}{G}
\end{eqnarray}
The two bunches of vertical lines represent
the unperturbed energy of a pair of particles placed in a given potential. The GPV 
is the collective state relative to the second major shell. It is analogous to the giant resonances of nuclear 
shapes which involve the coherent superposition of ph intrinsic excitations. 
 %uncomment the following lines to place a figure

As in the case of the low-lying PV,  the GPV 
should be populated through (p,t) or (t,p) reactions but despite  
efforts so far it has never been identified~\cite{Crawley,Mouginot,Zyb87}. 

Very recently Refs. \cite{Cappuzzello,Cavallaro1} reported on experiments to investigate
the GPV mode in light nuclei, using heavy-ion-induced two-neutron transfer reactions. 
The reactions $^{12}$C($^{18}$O,$^{16}$O)$^{14}$C  and $^{13}$C($^{18}$O,$^{16}$O)$^{15}$C 
 were studied at 84 MeV incident laboratory energy.  
"Bump-like" structures in the excitation energy spectra were identified as the GPV states in
$^{14}$C   and $^{15}$C  nuclei at excitation energies of  $\approx$ 20 MeV.  
Their energies and L=0 nature, as well as the extracted transfer
probabilities are consistent with
the GPV population.  It still remains as an intriguing puzzle that this mode has not been observed in heavier systems like Sn and Pb isotopes, where the collective effects are expected to be much stronger and for which the low-lying pair excitations are well described by pairing rotations in the Sn's and by pairing fluctuations near the critical point in the Pb's  \cite{BB,CMFK}.

The goal of this work is to provide an overview of both theoretical and experimental studies of this collective pairing mode in heavy nuclei.  The article is organized as follows:   In Section II we give an overview of the theory of the GPV;   in Section III we review the status of experimental searches, and  in Section IV we discuss the effects of Q-value mismatch on the cross-sections and the anticipated advantage of using weakly bound projectiles.  We conclude our manuscript by addressing 
some open questions and speculating on possible future studies. 
 
The observation and characterization of the GPV in light nuclei is being discussed in detail by Cavallaro et al. in this issue of EPJA~\cite{Cavallaro}.

\section{ Overview of the theory of the GPV }
%\label{sec:1}
%and \cite{RefJ}
Collective excitations in nuclei were recognized at the very beginning of 
nuclear studies with the introduction of the liquid drop model by Gamow in
1930 \cite{gam30}. This model had a great success, particularly in the study 
of nuclear masses \cite{wei35}, neutron
capture \cite{boh36} and nuclear fission \cite{boh39}. But the influence of
the liquid drop model went beyond these outstanding applications. One
can assert that both the rotational and the vibrational models were inspired
by the liquid drop model. From the viewpoint of this paper the important
outcome of the liquid model is the vibrational motion. Before the
appearance of the shell model vibrations were envisioned as a macroscopic
motion of the nucleus vibrating along a stable liquid drop surface. The
first microscopic calculations of surface 
vibrational states were performed in terms
of particle-hole excitations using a harmonic oscillator representation and
a separable force. One thus obtained the collective vibration as the lowest 
correlated particle-hole state. Here collective means that the strength of
the electromagnetic transition probability, $B(E\lambda)$,  is large due to the
coherent contribution of all particle-hole configurations  
\cite{bro59}. Soon afterwards it was found that including high lying shell
model configurations the $B(E1)$ strength corresponding to dipole excitations 
concentrates in the uppermost state
(instead of in the lowest one mentioned above). This state, which accounts for 
most of the electromagnetic energy waited sum rule, is the giant dipole resonance
\cite{bro60}.

This  type of shell model calculations was enlarged by including correlations in the ground state.
This was performed in the framework of the particle-hole Green function. Using the ladder approximation this treatment 
 turned out to be equivalent to the RPA
\cite{bro61}. One thus obtains the correlated (vibrational) gound state, with
hole-particle (backward RPA) configurations which in some cases are rather
large. One important feature of the RPA treatment is that there
is a value of the strength of the interaction large enough which
yields complex values for the energies, i. e. the RPA eigenvalues. At this
point there is a phase transition in the nucleus from a spherical to a deformed
shape \cite{bro71}.

There is a formal equivalence between particle-hole and two-particle
excitations \cite{BHR}.  Studies performed within the 
two-particle Green function using the ladder approximation lead to 
the two-particle RPA that give eigenvalues corresponding to states
of both A+2 and A-2 systems, where A is the nucleon number in the spherical 
normal core \cite{mul67}. As in the particle-hole case, it is found that 
if the interaction strength is large enough then there is a phase transition, in this 
case from a normal to a superconducting state.

The collective character of the particle-hole (surface) vibration is probed by
inelastic scattering reactions. In the same fashion two-particle transfer 
reactions provide much  of  our knowledge of pairing correlations. 
For excitations to $0^+$ states these reactions are 
important probes of collective pairing excitations in nuclei. This has the same origin
as the collectivity of surface vibrations in inelastic
scattering. Namely all configurations contribute with the same phase to the two-particle transfer form factor leading
to the collective pairing state (which can be considered a vibration
in gauge space \cite{BM}).  As we will show later,  
the cross section corresponding to pairing vibrations is much 
larger than those corresponding to other $0^+$ states.

The analogy between the surface and pairing modes goes even farther. In Ref.
\cite{BrogliaBes} it was predicted that a collective pairing vibration induced by excitations
of pairs of particles and holes across major shells should exist at an energy
of $\approx$ 65/\rm A$^{1/3}$ MeV carrying a cross section which is 20\% --100\% of the
ground state cross section. However, it is important to point out in the context of this paper that the (absolute) cross section leading to the GPV as predicted above is not as large as the one leading to particle-hole giant resonances in inelastic scattering.

Since the GPV was not observed experimentally, the subject gradually 
lost its interest from a theoretical perspective.  However, it revived independently and in a completely 
different framework more than a decade later,  in relation to alpha-decay as we discuss below.  
   
One standing problem in alpha-decay is the evaluation of the absolute decay
width, i. e. of the half life of the decaying state. The decay process takes
place in two steps. First the alpha-particle is formed on the surface of the
mother nucleus and in a second step the alpha-particle thus formed
penetrates the centrifugal and Coulomb barriers. The calculation corresponding
to this second step is relatively easy to perform since it is  just the penetrability introduced by Gamow
in his seminal paper of 1928 \cite{gam28}. The great difficulty is to
evaluate the alpha formation probability. In the beginning one expected that
this calculation was feasible  within the framework of the shell model, since 
the shell model is, more than a model, a tool that provides an excellent
representation to describe nuclear properties \cite{ras65}. In the first calculation
only one shell-model configuration was used \cite{man60} 
due to the inadequate computing facilities at that time. The results were discouraging 
since the theoretical decay rates were smaller than the corresponding 
experimental values by many orders of magnitude. It was eventually found that
the reason of this huge discrepancy was due to the lack of configurations in
the shell model basis \cite{ton79}. It was soon realized that the physical
feature behind the configuration mixing was that the clustering of the two
neutrons and two protons that eventually become the alpha particle proceeds
through the high-lying configurations \cite{jan83}. That was shown in
spherical normal systems, but even in deformed and superfluid nuclei that 
property is valid \cite{cat84}. Moreover, it was also
found that the same feature holds for the particle and the hole that
constitute the surface vibration \cite{fer88}. 

Calculations performed to evaluate the half life of the ground state of  $^{212}$Po, with two neutrons and 
two protons outside the $^{208}$Pb core, including a large number of neutron-neutron and proton-proton
configurations were still in disagreement with experimental data by about one order of magnitude \cite{ton79,jan83}.
It was realized that this disagreement was due to the lack of any neutron-proton interaction. That is, that calculation
included the neutron-neutron and proton-proton clusterizations but not the neutron-proton one. This was done
in Ref. \cite{dod85}, where the wave function of  $^{212}$Po(gs) was assumed to be
\begin{eqnarray}
\rm{|^{212}Po(gs)> = A  |^{210}Pb(gs)\otimes^{210}Po(gs)>} \nonumber \\
\rm{+ B|^{210}Bi(0^+_1)\otimes^{210}Bi(0^+_1)>}
\end{eqnarray} 
where A and B are constants to be determined (for details see Ref. \cite{dod85}). 
One sees in this equation that the first term corresponds to the clustering of the neutrons through the isovector 
pairing state $\rm ^{210}Pb(gs)$, with $T=1,T_z=1$ and the protons through the isovector 
pairing state $\rm ^{210}Po(gs)$, with $T=1,T_z=-1$ while the second term corresponds to the neutron-proton clustering
through the isovector pairing state $\rm ^{210}Bi(0^+_1)$, with $T=1,T_z=0$. But this last state was not measured at that time (and it is not at present either). Since in $\rm ^{210}Bi(0^+_1)$ both neutrons and protons move in the shells above N=126, which implies that the proton moves in an excited major shell,
it was assumed that this $T_z=0$  pairing state  lies at 5 MeV, which is about the energy 
difference between two major shells in the lead region. Using appropriate values of A and B one found a very strong clustering of the four nucleons that constitute the alpha-particle, and a  $\rm ^{212}$Po(gs) half life which was in perfect agreement with experiment. This showed the importance of the neutron-proton clustering, but unfortunately the values of A and B thus used were unrealistic. 

The feature that has to be underlined for the purpose of this paper is that the states 
$\rm ^{210}Pb(gs)$ and  $\rm ^{210}Bi(0^+_1)$ are isoanalogous, and that the third component of these three T=1 states, 
with $T=1, T_z=-1$ should be a  collective pairing state lying at about 10 MeV. This is the state  $\rm ^{210}Po(0^+_{GPV})$. Thus in the lead isotope the equivalent of 
$\rm ^{210}Bi(0^+_1)$ is $\rm ^{210}Pb(gs)$. But there should also exist the state $\rm ^{210}Pb(0^+_{GPV})$. Unaware of the work of 
Ref. \cite{BrogliaBes}, in Ref. \cite{her86} one  evaluated again $\rm ^{210}Pb(0^+_{GPV})$ finding a large neutron-neutron clustering  and a large two-neutron cross section leading to the GPV. 

\section{ Status of experimental searches for the GPV}

\label{sec:3}

\subsection{ The population of the GPV in two-nucleon transfer reactions }

As discussed in the previous  Section, an important consideration in the observability of the GPV is the coherence in the mixed wave functions.  This  is  expected to enhance the observed cross sections as the different amplitudes for the two-particle transfer operator have the same sign and add coherently~\cite{Oer01}.  
As a measure of the collectivity, we then look at the transfer operator, realizing that a realistic estimate should take into account the kinematic features of the two nucleon transfer cross sections to 0$^+ $ states, by considering a DWBA calculation.  The 2-nucleon transfer operator plays a similar role to the $B(E\lambda)$ for surface modes.

Given a set of single particle orbits  $\ket{n \ell j} \equiv \ket{j} $, the wave function of the GPV state can be written:

\begin{align}
\ket{GPV} =  \sum_{j}\alpha_j\ket{j^2}  ; \nonumber 
\end{align}
The matrix element for the transfer of a pair of L=0 neutrons to the GPV  in nucleus $|A_0+2>$ from the ground state of $|A_0>$ is
\begin{align}
\bra{GPV} T \ket{A_0} =  \sum_{j}\alpha_j  \bra{j^2} T \ket{0}  ; \nonumber 
\end{align}
and the cross section 
\begin{align}
\sigma(GPV)  \propto \bra{GPV} T \ket{A_0}^2 =   ( \sum_{j} \alpha_j)^2  \sigma_{sp}; \nonumber 
\end{align}
with the further assumption that the single particle matrix elements are all approximately equal,  $ \bra{j^2} T \ket{0}^2 \approx \sigma_{sp}$.  As we will discuss later, this simplification is not always realistic.

The limiting case of $\Omega$ degenerate levels provide an estimate of the maximum collectivity.  Here we have $\alpha_j \approx \frac{1}{\sqrt{\Omega}}$
and thus
\begin{align}
\sigma(GPV)  \sim \Omega \sigma_{sp} 
\end{align}
which should scale with mass number as $\sim \rm A^{2/3}$. A realistic example of the enhancement in the population of  collective pairing modes is illustrated in Fig. 2,  comparing the pairing strength $ \bra{GPV} T \ket{A_0}$ for the addition modes in
$^{208}$Pb calculated in the Tamm-Dancoff (TDA) and RPA approximations with the unperturbed results. 

\begin{figure}
\resizebox{0.45\textwidth}{!}{ \includegraphics{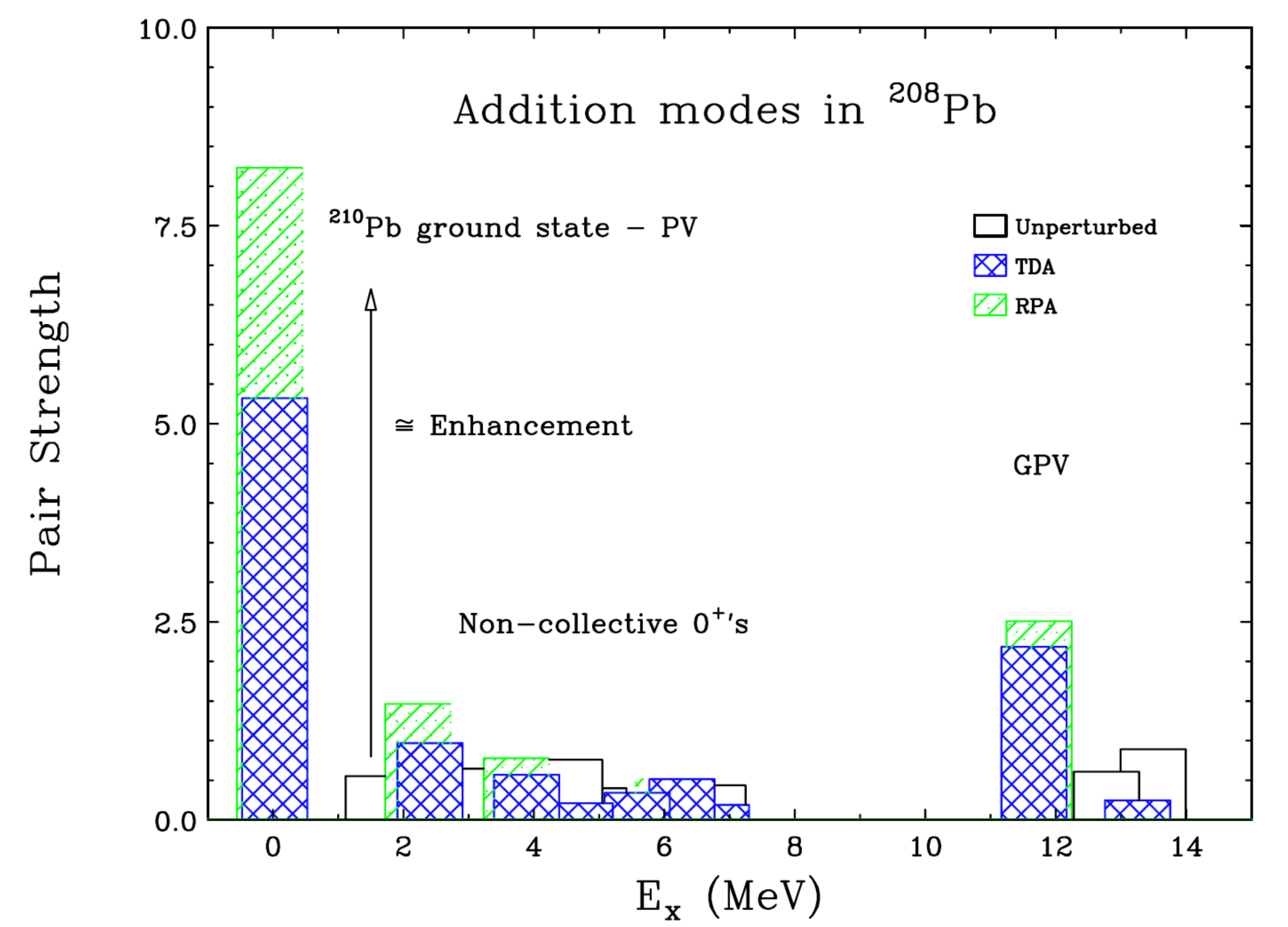}}
\caption{(Color online) TDA and RPA results for the pair strength for the addition of two neutrons on the gs of $^{208}$Pb. The enhancement (with respect to the unperturbed results) in the population of the PV and the GPV is clearly seen. }
\end{figure}

\subsection{Search for GPV through (p,t) reactions}
\label{subsec:3.1}

The simple estimate in Eq. (4) shows that the collectivity of the GPV increases with the mass of the nucleus. Therefore,  the pair strength is expected to be maximum for the heaviest nuclei, such as Sn and Pb isotopes, where numerous nucleons may contribute coherently. Two candidate regions of the nuclear chart have been envisaged:  in Pb  (€closed-shell, normal nuclei)  and the Sn's (mid-shell, superfluid nuclei). In these nuclei, the GPV is supposed to be typically located around 12 MeV and 14 MeV respectively.  So far, the GPV in those nuclei has not been found, although a great experimental effort was devoted to it using (p,t) and (t,p) reactions in various conditions.

In the 60's and 70'€˜s, the searches for the GPV focused on (p,t) reactions at high energy for both Pb and Sn isotopes. However they remained unsuccessful. There could be several reasons as mentionned in \cite{Mouginot} :
\begin{itemize}
\item The L matching conditions are an of great importance. The proton incident energy should be high enough to excite a 14 MeV mode but not too high in order not to hinder the L=0 transfer. The smaller the proton energy the larger the cross section for L=0 modes.
\item The use of a spectrometer is decisive in order to precisely measure the triton in the exit channel. The only reported search for the GPV with Ep $\approx$ 50 MeV used Si detectors, and was plagued by a strong background \cite{Crawley}.
\item As the L = 0 cross sections are known to exponentially increase when approaching 0 degree, the measurement has to be performed at small angles and is even better if it includes 0 degree.
\end{itemize}

% Experiments @ iThemba
There was a revival of the experimental GPV searches in the 2000's with several experiments aiming at improving the three experimental conditions mentioned above. All used a spectrometer for the triton measurement to improve the measurement at 0 degree. Several attempts with different proton energies were performed.
The first attempt used a 60 MeV proton beam produced at the iThemba LABS facility in South Africa impinging on $^{208}$Pb and $^{120}$Sn targets \cite{Mouginot}. The tritons were measured at 7 degrees with the K = 600 QDD magnetic spectrometer. The strong deuteron background was removed thanks to their different optical characteristics. No evidence for the GPV was found in the region of interest in neither of both targets.

\begin{figure}
\resizebox{0.5\textwidth}{!}{ \includegraphics{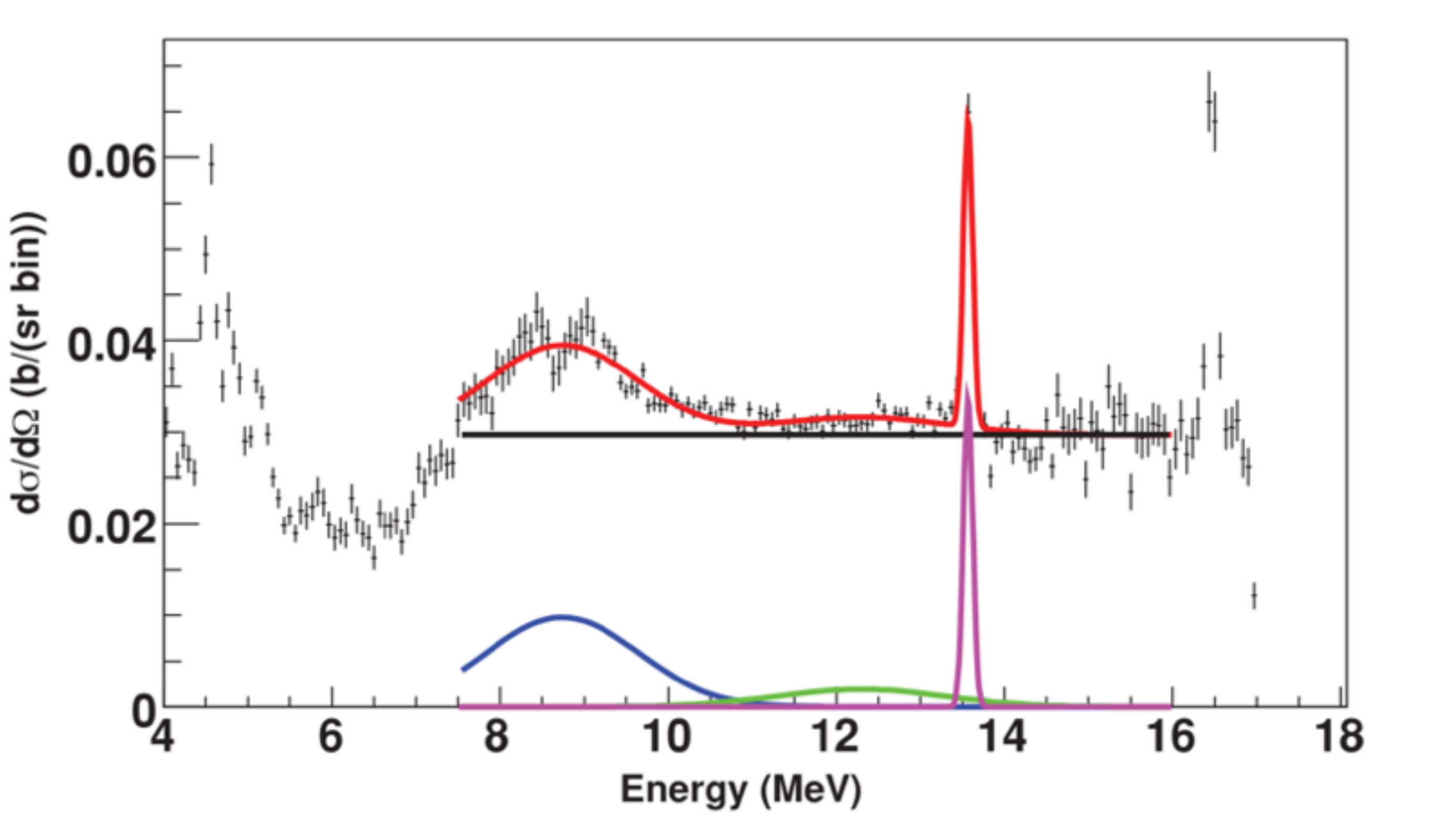}}
\caption{(Color online)  Excitation energy spectrum of $^{118}$Sn for the 0 degree measurement at Ep = 50 MeV. The individual fits for a linear background (black), deep-hole states (blue), a possible GPV around 12 MeV (green), and  oxygen contaminant (magenta) are shown together with the total fitting function (red). The bin width is 67 keV/bin. From  \cite{Mouginot}.\label{fig_GPV_iThemba}}
\end{figure}

The measurement was repeated with 50 MeV and a 60 MeV proton beams and the K = 600 QDD magnetic spectrometer in zero degree mode to combine the best experimental conditions to probe the GPV. In this case the beam stopper, placed midway between the two dipole magnets of the spectrometer, produced a strong proton background with a rate $\sim$ 500 times higher than that of the tritons of interest. This background consisted of protons scattering off the beam stop with the combinations of angles and magnetic rigidities so that their trajectories reached the focal plane detectors. The time of flight between the SSC radio-frequency (RF) signal and the scintillator (from the spectrometer focal plane detection) trigger lead to the triton identification and removed most of the background. The excitation energy spectrum obtained for $^{118}$Sn is shown in Fig. \ref{fig_GPV_iThemba}. The deep holes contribution be- tween 8 and 10 MeV is stronger in the 0 degree spectrum than at 7 degrees indicating a possible low L composition of this region of the spectrum.  Assuming a linear dependence, obtained by averaging the background between 14 and 16 MeV, a fit of the different components assuming a width between 600 keV and 1 MeV for the GPV was performed. It leads to a higher limit on the cross-section for populating the GPV between 0.13 and 0.19 mb over the angular acceptance of the spectrometer ($\pm$2 degrees).

% Experiment @ Catania

\begin{figure}
\resizebox{0.5\textwidth}{!}{ \includegraphics{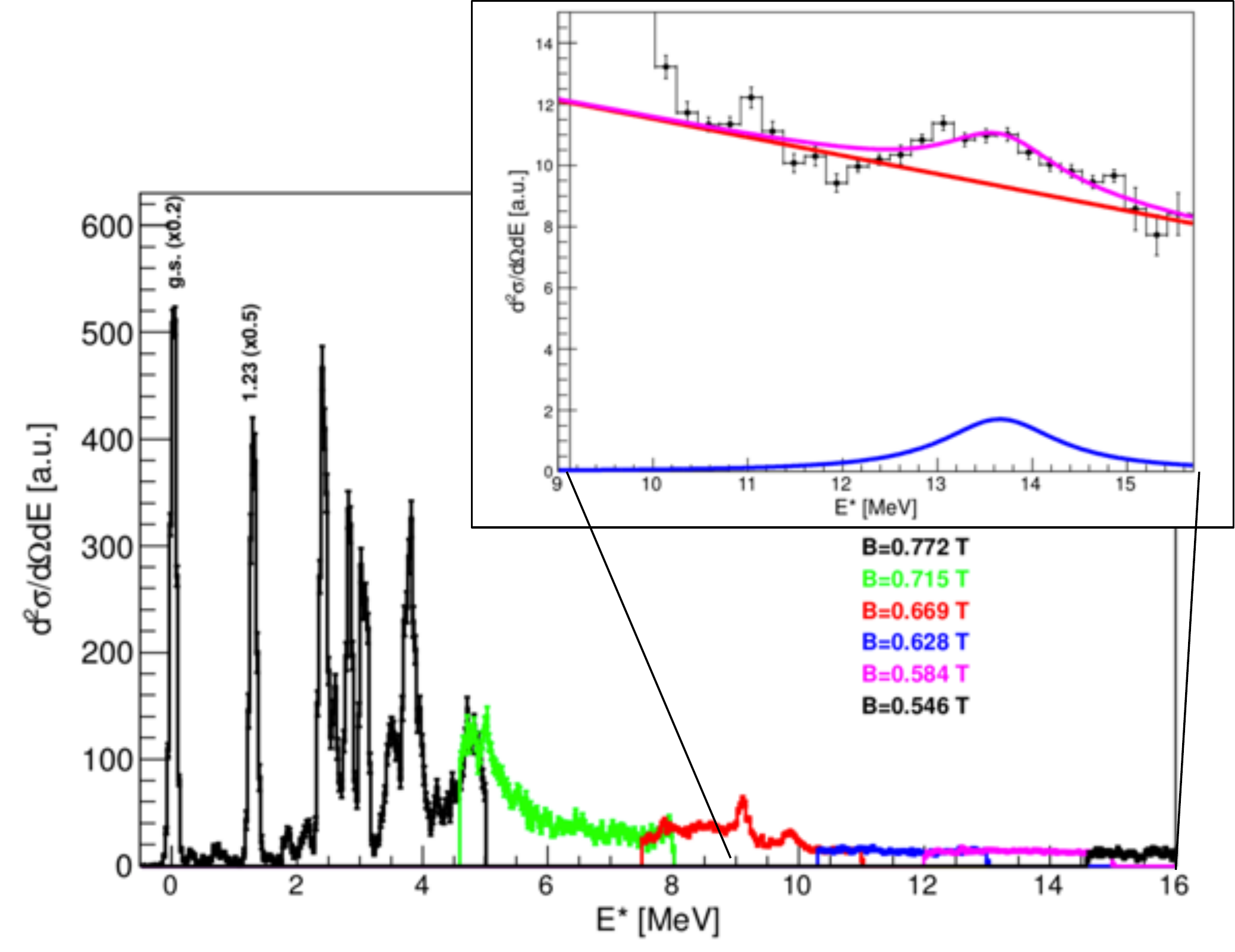}}
\caption{(Color online)  (Color online) Excitation energy spectrum of $^{118}$Sn for the six spectrometer settings at Ep = 35 MeV. A zoom on the GPV region is shown in the insert where a lorentzian fit of the GPV was performed (blue). From  \cite{deN14}.\label{fig_GPV_Catania}}
\end{figure}

The last attempt with the (p,t) reaction was performed at LNS Catania with a proton beam produced by the cyclotron accelerator at Ep = 35 MeV impinging on a $^{120}$Sn target \cite{deN14}. The lower proton energy was supposed to enhance the L=0 cross-sections and favor the population of the GPV. The measurement was performed with the MAGNEX large acceptance spectrometer. Tritons with energies between 12 and 18 MeV are expected for a GPV between 10 and 16 MeV. The MAGNEX energy acceptance is $\pm$25\%, which allows to cover a range of about 7 MeV in the expected GPV energy region. The excitation energy function obtained for $^{118}$Sn is shown in Fig.\ref{fig_GPV_Catania} for the six magnetic settings of the spectrometer. The tritons were identified from their energy loss as a function of their position in the focal plane so that a very small background contribution remains. The spectrum zoomed in the region of interest for the GPV shows a small bump over the background in the same energy region as the previous measurements at 50 and 60 MeV. The width was fitted to 1.5 $\pm$ 0.4 MeV. No clear evidence for a GPV mode has been found from the searches through (p,t) reactions. Improved experiments with (t,p) transfer reactions should  be revisited to rule out any difference between two-neutron stripping and two-neutron pick-up reactions.

\section{ Q-value effects on the reaction cross sections}

 As it is well known from the theory of 2-nucleon transfer reactions, there is an optimum Q-window for the transfer to occur \cite{Oer01,BrogliaWinter}.
The two-particle form factor, entering in the cross-section calculations, includes the overlap between the distorted scattering waves in the entrance and exit channels. If these waves are very different then that overlap is small, and the cross section itself will be small. A measure of the differences between those scattering waves is the reaction Q-value. A large Q-value means small overlap.    This translates into an exponential dependence quenching the cross-section outside the optimum Q-value, 
\begin{align}
\sigma \sim \rm exp \big[- \frac{(\rm Q - \rm Q_{opt})^2}{2 \hbar^2 \kappa \ddot{r}_0}\big]
\end{align}
where $\kappa$ is the slope of the two-particle transfer effective form factor and $\ddot{r}_0$ the acceleration at the distance of closest approach $r_0$.
Thus,  a possible reason to explain why the GVP has not been seen experimentally relies on the fact that both (p,t) and (t,p) reactions are well matched for gs to gs transitions, but the large excitation energy of the GPV hinders the cross-section more than it is enhanced by the coherence in the wavefunction.
Refs.~\cite{Fortunato,Dasso} have studied in detail the problem of exciting high-energy collective pairing modes in two-neutron transfer reactions.  Relaying on the analogy with the surface modes,  they used a collective form factor

\begin{align}
(\frac{\beta_P}{3A})R_0\frac{\partial U(r)}{\partial r}
\end{align}
with $\beta_P$ the deformation parameter of the pairing field\footnote{Typically given by $ \approx  \frac{\Delta}{G}$, the ratio of the pairing gap $\Delta$ to the strength of the paring force $G$, a measure of the available levels for scattering the pairs $\Omega$.} ~\cite{DV}, as input to the DWBA calculations. The results confirmed 
that, using conventional reactions with standard beams, one is faced with a large energy mismatch that favors the transition to
the ground state over the population of the high-lying states. Instead, the Q-values in a
stripping reaction involving weakly bound  nuclei are much closer to the
optimum for the transition to excited states in the 10-15 MeV range. 

\begin{figure}
\resizebox{0.5\textwidth}{!}{ \includegraphics{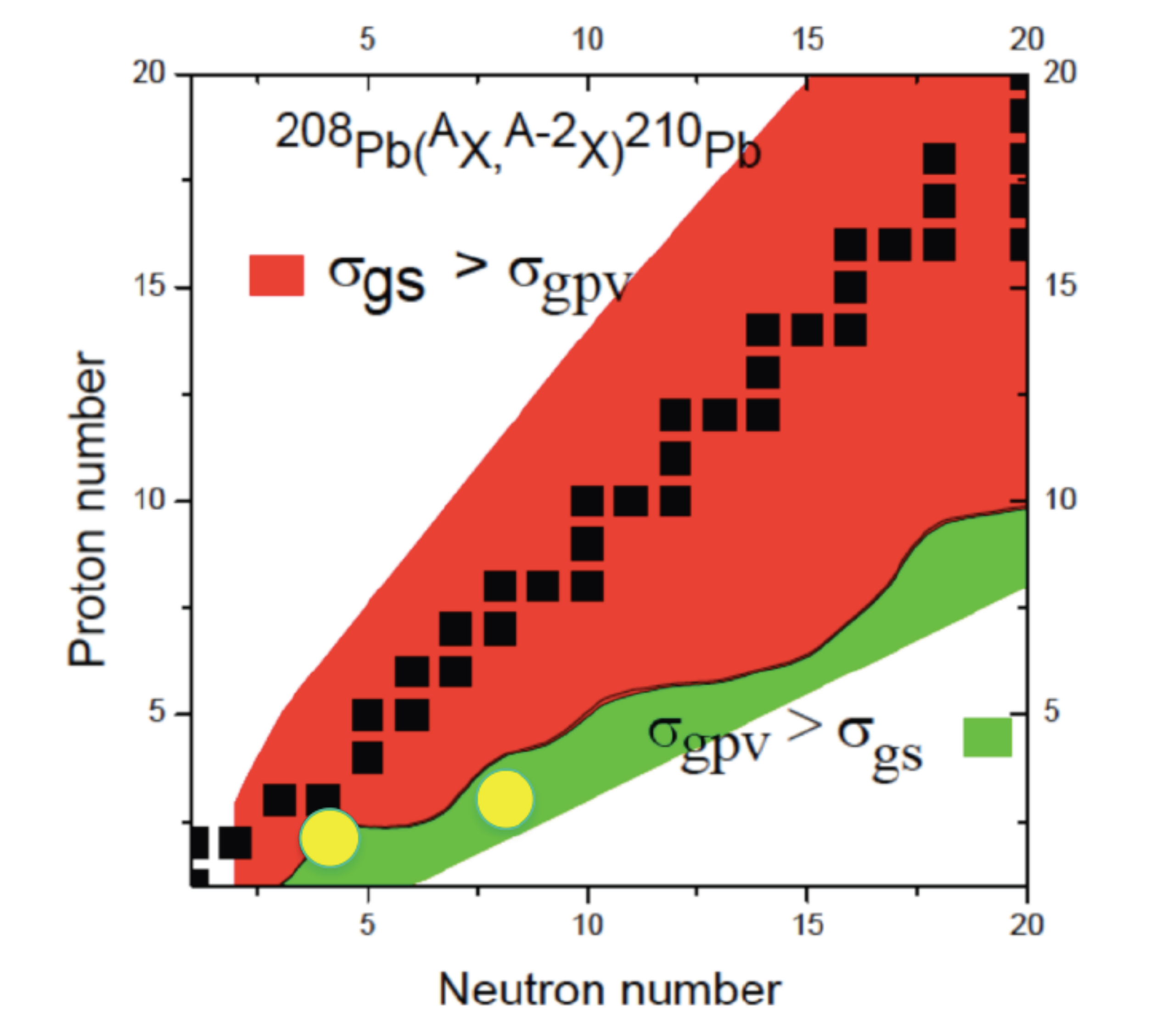}}
\caption{(Color online)  (Color online) Survey map for the $^{208}$Pb($^A$X,$^{A-2}$X)$^{210}$Pb reaction indicating the cases for which the transfer cross-section to the GPV 
is larger than to the ground state (green) Ref.~\cite{Dasso}.   $^6$He and $^{11}$Li beams are indicated with yellow circles.}
\end{figure}

Fig. 4 shows a survey map for possible projectiles ( $\rm ^AX$), for which the cross-sections to populate the GPV in $^{208}$Pb are anticipated to be larger than that to the 
gs.  An inspection of the Figure suggests the use of $^6$He and $^{11}$Li beams.   Transfer strengths and cross-section are compared to the case of $^{18}$O induced 
reactions in Fig. 5. The effect on the cross-section due to the Q-value mismatch is clearly seen.

%\subsection{ Using weakly bound projectiles }

\subsection{  Search for GPV through ($^{6}$He,$^{4}$He) reactions}
\label{subsec:3.2}
% Experiment @ GANIL
Following from the discussions above,  the $^{208}$Pb($^{6}$He,$\alpha$) reaction has been investigated at GANIL~\cite{Ass09} with the $^{6}$He beam produced by the Spiral1 facility at 20 MeV/A with an intensity of 10$^{7}$ pps. The detection system was composed of an annular Silicon detector. The background due to the various channels of two-neutron emission from $^{6}$He into $^4$He +2n and also to the channeling in the detector of the elastically scattered $^{6}$He beam was large 
and no indication of the GPV was found in this experiment.  The results of Ref. \cite{data} show a similar situation.  More recently, the reaction $^{116}$Sn($^{6}$He,$\alpha$) at 8 MeV/A was studied at TRIUMF~\cite{triumf} with the IRIS Array~\cite{Ritu}.  The analysis of these data is still in progress but due to the breakup background is too early to make any conclusions.

\begin{figure*}
\resizebox{1.0\textwidth}{!}{ \includegraphics{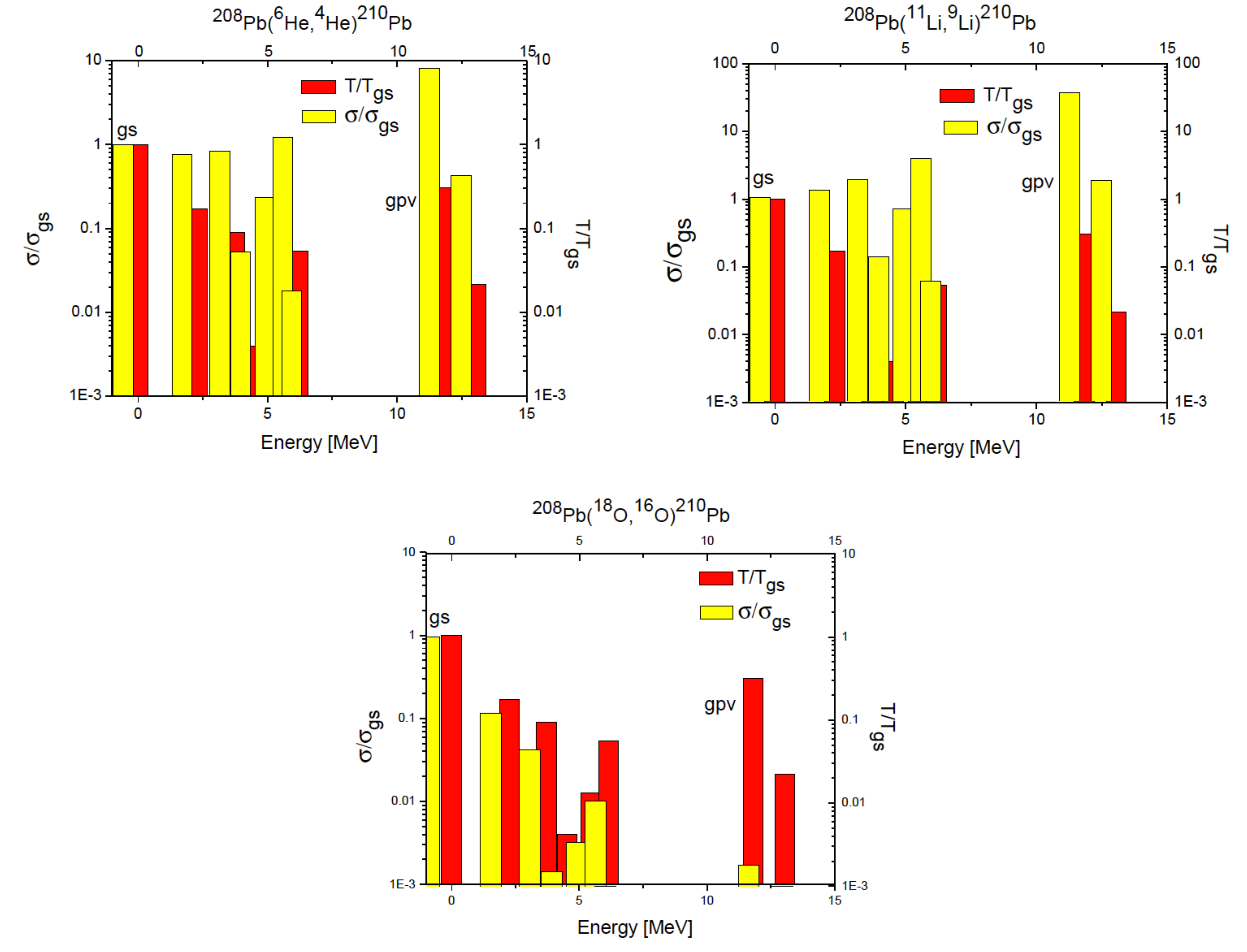}}
\caption{(Color online)  (Color online)Pair addition strength distribution, normalized to the ground state, to all $0^+$ states
in $^{210}$Pb (red bars) compared to the normalized pair transfer cross section (yellow bars) in
the case of $^{6}$He (upper left), $^{11}$Li ( upper right) and $^{18}$O (lower frame). From Ref.~\cite{Dasso}
 }
\end{figure*}

\label{sec:5}

\section{ Open questions  }

\subsection{The 2n-transfer form factor and cross-sections}

As discussed in Ref.~\cite{Laskin},  the two-nucleon transfer cross sections to $0^+$ states depend not only on the coherence of the wave functions but on the specific amplitudes for transfer of angular momentum zero-coupled pairs for different single-particle states entering in the forma factor. Basically, this reflects the probability for finding a $^1S_0$ 2n pair in the configurations $ \ket{(n\ell j)^2,L=0}$ \cite{Bayman}. These amplitudes depend strongly on the orbital angular momentum $\ell$, and the transfer probability could drop by order(s) of magnitude for each increase $\Delta \ell = 2$. Hence the bare cross section at the first maximum of the angular distributions for, say, two nucleons in an $i_{13/2}$ orbit, will be about 4 orders of magnitude less than that for the transfer of two $s_{1/2}$ particles. This effect is likely to be more important in the final cross sections than the detailed collectivity of the final states.    The selectivity of different two-particle transfer reactions, such as (t,p), ($^{18}$O,$^{16}$O), and ($^{14}$C,$^{12}$C), with
respect to detailed microscopic configurations in initial and final target states has recently been investigated in Ref. \cite{Lay}.

\subsection{ Weak binding and continuum effects }

The influence of the continuum on the properties of the giant pairing resonances was also motivated  
by alpha decay studies. As stated in Ref. \cite{ver88}, it is necessary 
to include the continuum in e. g. the formation of the alpha-particle in alpha 
decay and in the building up of resonant states lying high in nuclear spectra. 
In Ref. \cite{ver88} a representation was used consisting of bound states,
resonances and the proper continuum (composed by scattering waves) in the 
complex energy plane. This is the Berggren representation \cite{ber68}. In
this representation the scalar product between two vectors, i. e. the
metric, is the product of one vector times the other (instead of the complex conjugate of the 
other). But this affects only the radial 
part of the wave functions. The angular and spin parts are treated as usual. 
Since the radial parts of the wave functions can be chosen to be real 
quantities (e. g. harmonic oscillator functions for bound states, sinus and
cosinus functions for scattering states) the Berggren scalar product 
coincides with the Hilbert one on the real energy axis. Therefore the space 
spanned by the Berggren representaion (the Berggren space) can be considered a generalization of the 
Hilbert space. It has been shown that this representation is indeed a
representation, that is that it can describe any process in the complex
energy plane \cite{lio96}.

Within the Berggren representation the energies may be complex. Gamow showed that in a time
independent context a resonance can be understood as having complex energy \cite{gam28}.
The real part of this energy corresponds to the position of the resonance
and the imaginary part is, in absolute value, half the width. The resonances
entering in the Berggren representation are Gamow resonances.
Due to the metric of the Berggren space, not only energies but also
transition probabilities related to the evaluated states can be complex. 
A many-particle state lying on the complex energy plane may be considered a resonance, i.e. a
measurable state appearing in the continuum part of the spectrum, if the
wave function is localized within the nuclear system. This usually happens if the imaginary 
part of the energy (i. e. the width) is small \cite{bet05}. Otherwise the state is just a part of
the continuum background. This property will be important in the analysis of the giant pairing vibration.

By using the Berggren representation the shell model was extended to the complex energy
plane given rise to the complex shell model and the Gamow shell model. A review on this can be found in \cite{mic09}. 

The Bergen representation was used to analyze particle-hole resonances within a RPA formalism \cite{cur89}. It was thus found that in $\rm ^{208}Pb$ the escaping widths of the giant resonances, which lie well above the neutron escape threshold,  are small because the particle moves on bound shells or narrow Gamow resonances, while the hole states are all bound. But from the viewpoint of this paper the important outcome of calculations in the complex energy plane was the study of giant pairing vibrations performed in Ref. \cite{dus09}. Before this, one used bound (e. g. harmonic oscillator) representations, which did no consider the decaying nature of the resonances. Instead, it was found that within the Berggren representation the two-neutron GPV in $\rm ^{210}Pb$ is very wide and is not a physically relevant state but a part of the continuum background. The proton-neutron GPV in $\rm ^{210}Bi$ was found to be a meaningful state only if it is not a resonance but a bound state lying below 7 MeV of excitation. As this energy approaches the continuum threshold, then the collectivity of the state gradually disappears. Above the threshold not only does the collectivity vanish altogether but also the resulting resonance is very wide. For details see \cite{dus09}.

In Ref.~\cite{Laskin} the formalism developed by von Brentano, Weidenmuller and collaborators for mixing of bound and unbound levels~\cite{Brentano,Brentano2} was applied to 
the study of simple toy-model and realistic calculations to asses the effects of weak binding and continuum coupling on the non-observation of the GPV.   It was found that the mixing
in the presence of weak binding was a minor contributor to the weak population.  Rather, the main reason was attributed to the melting of the GPV peak due to the width it acquires from the low orbital angular momentum single particle states playing a dominant role in two-nucleon transfer amplitudes. This effect, in addition to the Q-value mismatch, may account for the elusive nature of this mode in (t, p) and (p, t) reactions.

In summary,  the continuum part of the nuclear spectrum appears to be more important for the pairing vibration mode than for the surface vibration one. This is because the two particles in the pairing mode at high energy escape easily into the continuum. Instead the only particle lying in the continuum in the particle-hole mode moves largely in narrow resonances, thus been trapped within the nucleus during a time long  enough for the resonance to be seen. Citing Bortignon and Broglia~\cite{PFB}, {\sl "... the fact that the GPV have likely been serendipitously observed in
these light nuclei when it has failed to show up in more propitious nuclei like Pb, provides unexpected and
fundamental insight into the relation existing between basic mechanisms -- Landau, doorway, compound
damping -- through which giant resonances acquire a finite lifetime, let alone the radical difference regarding
these phenomena displayed by correlated (ph) and (pp) modes."}

\section{Future studies and conclusions}

in spite of several experimental efforts, the elusive nature of the GPV in heavy nuclei remains as an intriguing puzzle.  Severe Q-value quenching of the cross-sections for (t,p) and (p,t) reactions has suggested the use of weakly bound projectiles,
such as $^6$He and $^{11}$Li, to overcome those limitations.  Unfortunately the large 2n breakup probability conspires to mask the GPV signal with a large background. Nevertheless,
further exclusive measurements should be carried-out in order to either rule out the population  of the GPV or establish a firm limit that could be compared to theory.

The availability of state-of-the art instrumentation, tritium targets, and possibly tritium beams also suggests that the (t,p) reaction should be revisited.  
Furthermore, since from the point of view of the continuum effects both the proton-neutron GPV in $\rm ^{210}Bi$ and the proton-proton GPV in $\rm ^{210}Po$ are anticipated to be meaningful states,  a search for these resonances using for example the ($^{3}$He,p) and ($^{3}$He,n) reactions should be pursued.  One could even speculate on using the $(\alpha,$d) reaction combining particle and gamma-spectroscopy to tag on the $^2$H 2.2 MeV gamma to select the transfer of an isovector np pair. 

Another independent way of probing the GPV is by exploiting the $T = 1$ isobaric character of the states $^{210}$Po(GPV),
$^{210}$Bi(GPV), and $^{210}$Pb(gs).   By means of charge-exchange reactions like (p,n) or ($^3$He,t) on a $^{210}$Pb (radioactive) target or in inverse kinematics using a $^{210}$Pb (radioactive) beam to populate the $^{210}$Bi(GPV).

Finally, given the fact that recent theoretical efforts have pointed out the important effects of weak-binding and continuum coupling, realistic estimates of the total damping width (Escape width, Landau and doorway damping~\cite{GRbook}) of the GPV in Sn and Pd isotopes will be extremely valuable.  

To conclude, after more than fifty years since the analogy between atomic nuclei and the superconducting state in metals was pointed out in Ref.~\cite{BMP}, the role of  pairing correlations in nuclear structure continues to be a topic of much interest and excitement in the field of nuclear physics~\cite{BZ}.  The discovery of the GPV in light nuclei opens up a
unique opportunity to advance our knowledge of high-lying pairing resonances but, at the same time,  the non-observation of these modes in heavy nuclei remains as an open question that needs to be further addressed both from theory and experiment.

\section{Acknowledgments}
This material is based upon work supported by the U.S. Department of Energy, Office of Science, Office of Nuclear Physics under Contract No. DE-AC02-05CH11231 (LBNL). One of us (AOM) would like to thank Rod Clark and Rick Casten for many discussions on the topic.
%\subsection{ Damping width    \textcolor{red}{Andrea} }
%\subsection{ NP GPV}
%\subsection{GPV in mesoscopic systems ...}

%
%
% BibTeX users please use
% \bibliographystyle{}
% \bibliography{}

\end{document}